\documentclass[aps,prl,twocolumn,superscriptaddress,footinbib]{revtex4-1}
\usepackage{amsmath,amssymb,bm}
\usepackage{graphicx}
\usepackage{epstopdf}
\usepackage{latexsym}
\usepackage[normalem]{ulem} 
\usepackage[usenames,dvipsnames]{color}
\usepackage{hyperref}
\usepackage{natbib}
\usepackage{braket}
\begin{document}

\newcommand{\Rev}[1]{{\color{black}{#1}\normalcolor}} 
\newcommand{\Com}[1]{{\color{black}{#1}\normalcolor}} 

\title{Characterizing the dynamical phase diagram of the Dicke model via classical and quantum probes}

\date{\today}

\author{R.~J. Lewis-Swan}
\affiliation{Homer L. Dodge Department of Physics and Astronomy, The University of Oklahoma, Norman, Oklahoma 73019, USA}
\affiliation{Center for Quantum Research and Technology, The University of Oklahoma, Norman, Oklahoma 73019, USA}
\author{S.~R. Muleady}\thanks{These two authors contributed equally}
\affiliation{JILA, NIST, Department of Physics, University of Colorado,  Boulder, CO 80309, USA}
\affiliation{Center for Theory of Quantum Matter, University of Colorado, Boulder, CO 80309, USA}
\author{D. Barberena}\thanks{These two authors contributed equally}
\affiliation{JILA, NIST, Department of Physics, University of Colorado,  Boulder, CO 80309, USA}
\affiliation{Center for Theory of Quantum Matter, University of Colorado, Boulder, CO 80309, USA}
\author{J.~J. Bollinger}
\affiliation{National Institute of Standards and Technology, Boulder, Colorado 80305, USA}
\author{A.~M. Rey}
\affiliation{JILA, NIST, Department of Physics, University of Colorado,  Boulder, CO 80309, USA}
\affiliation{Center for Theory of Quantum Matter, University of Colorado, Boulder, CO 80309, USA}

\begin{abstract}
We theoretically study the dynamical phase diagram of the Dicke model in both classical and quantum limits using large, experimentally relevant system sizes. \Rev{Our analysis elucidates that the model features dynamical critical points that are strongly influenced by features of chaos and emergent integrability in the model.} Moreover, our numerical calculations demonstrate that mean-field features of the dynamics remain valid in the exact quantum dynamics, but we also find that in regimes where quantum effects dominate signatures of the dynamical phases and chaos can persist in purely quantum metrics such as entanglement and correlations. Our predictions can be verified in current quantum simulators of the Dicke model including arrays of trapped ions. 
\end{abstract}

\maketitle  

\noindent{\it Introduction:} Advances in atomic, molecular and optical (AMO) quantum simulators are driving a surge in the investigation of dynamical phase transitions (DPTs) and associated non-equilibrium phases of matter \cite{Heyl_DPTtheory_2013,Schiro10,Sciolla_2011,Halimeh_2017_prethermal,Halimeh_2017_longrange,Silva_DQPT_2018}. In a closed system, a DPT is a critical point separating distinct dynamical behaviours that emerge after a sudden quench of a control parameter \cite{lerose2018dpt,lerose2019dpt,schiro2011,Peronaci2015dpt,sciolla_2013,Chiocchetta_2015,Chiocchetta_2017}, and can be defined via non-analytic behaviour in a time-averaged order parameter \cite{eckstein09,gambassi,Knap,Lang_2018}.

To date, the majority of experimental investigations in this direction have been tailored towards integrable models, featuring effective infinite-range interactions between spins, which admit analytical treatments \cite{Thywissen_DPT_2018,Duan_DPT_2019,Duan_DPT_2020,Muniz2020}. Richer non-integrable models have been pursued in trapped ion systems  \cite{Monroe_DPT_2017}, but the associated complexity of the quantum dynamics limited the theoretical analysis of the DPT to small system sizes and prevented a rigorous scaling analysis. Hence, it is highly desirable to find and study DPTs in non-integrable models featuring novel non-equilibrium phenomena that are both implementable in tunable quantum simulators \emph{and} theoretically tractable under controllable approximations.

We advance this direction by studying a DPT in the iconic Dicke model \cite{Dicke1954,Garraway2011,Kirton_2019}, which describes the collective coupling of many spins to a single harmonic oscillator. The model is attractive as it features an array of phenomena, such as non-integrability \cite{Altland_2012,Altland_PRL_2012,Buijsman_2017}, chaos \cite{Chavez_2016,RLS_scrambling_2019,Villase_2020} and equilibrium quantum phase transitions (QPTs) in both ground and excited states \cite{Emary2003_PRL,Emary2003_PRE,Perez_2011,Brandes_2013,Puebla_2014,Hirsch_2014}, but involves only a pair of disparate degrees of freedom such that the model remains amenable to analytic and numerical treatments. Moreover, the model is already studied in state-of-the-art AMO quantum simulators, including trapped-ion arrays \cite{Bollinger_2018,Cohn_2018} and cavity-QED \cite{Baumann2010,Klinder2015,Barret_2018,Lev2018}. Here, we investigate the DPT in analytically tractable spin- and boson-dominated limits, as well as a non-integrable regime where near-resonant coupling of spin and bosons leads to chaotic dynamical phases also seen in other non-integrable systems \cite{lerose2018dpt,lerose2019dpt}.
\Rev{Following recent works linking DPTs to co-existing excited-state QPTs (EQPTs) \cite{Puebla_2014,Duan_DPT_2020,Puebla2020}, we find different dynamical critical points in the spin and boson-dominated regimes that reflect the presence of distinct EQPTs in these limits.}
By studying large, experimentally relevant, system sizes using efficient numerical methods we are able to demonstrate that mean-field features of the dynamics remain valid in the exact quantum dynamics. Conversely, in regimes where quantum effects dominate we find that signatures of the DPT and chaos persist in purely quantum metrics such as entanglement and correlations.

\noindent{\it Model:} The Dicke Hamiltonian for $N$ spin-$1/2$ particles is given by \cite{RLS_scrambling_2019},
\begin{equation}
    \hat{H}_{\mathrm{D}} = \frac{2g}{\sqrt{N}}\left( \hat{a} + \hat{a}^{\dagger} \right)\hat{S}_z + \delta\hat{a}^{\dagger}\hat{a} + \Omega\hat{S}_x .
\end{equation}
Here, $\hat{a}$ ($\hat{a}^{\dagger}$) is the bosonic annihilation (creation) operator of an oscillator with frequency $\delta$, $\hat{S}_{\alpha} =1/2 \sum_{j=1}^N \hat{\sigma}^{\alpha}_j$ are collective spin operators for $\alpha=x,y,z$ and $\hat{\sigma}^{\alpha}_j$ Pauli matrices of the $j$th spin. The spins are subject to a transverse field of strength $\Omega$ and the spin-boson coupling is characterized by $g$. The Hamiltonian exhibits a $\mathbb{Z}_2$ parity symmetry associated with the operator $\hat{\Pi} = e^{i\pi(\hat{S}_x + \hat{a}^{\dagger}\hat{a} + N/2)}$ such that $[\hat{H}_{\mathrm{D}},\hat{\Pi}]=0$. 

The equilibrium phase diagram of the Dicke model features a ground-state QPT at a critical coupling $g_{\mathrm{QPT}} = \sqrt{\delta\Omega}/2$ \cite{Emary2003_PRL,Emary2003_PRE}, which delineates a superadiant ($g \gg g_{\mathrm{QPT}}$) phase, with degenerate ground-states associated with different parity sectors \Rev{that in the limit $\Omega \to 0$ take the form} $\vert \psi_{s} \rangle = \frac{1}{\sqrt{2}}[\vert -(N/2)_z \rangle\otimes \vert \alpha_s \rangle \pm \vert (N/2)_z \rangle\otimes \vert -\alpha_s \rangle]$, and a normal ($g\ll g_{\mathrm{QPT}}$) phase, with $\vert \psi_n \rangle \approx \vert (N/2)_x \rangle\otimes \vert 0\rangle$ \Rev{in the limit of large $\Omega$} \cite{Bollinger_2018}. Here, we have defined collective spin states via $\hat{S}_{x,y,z}\vert m_{x,y,z} \rangle = m_{x,y,z}\vert m_{x,y,z} \rangle$ and $\vert \pm \alpha_s \rangle$ is the bosonic coherent state for $\alpha_s = g\sqrt{N}/\delta$. Additionally, an excited-state QPT (EQPT) exists in the spectrum of the superradiant phase, $g > g_{\mathrm{QPT}}$, defined by a critical energy $E_c = -\Omega N/2$ \cite{Perez_2011,Puebla_2014,Hirsch_2014}. The EQPT is signaled by a non-analyticity in the density of states \cite{Brandes_2013} and the existence of pairs of degenerate eigenstates with different parity for $E < E_c$.


\noindent{\it Dynamical phase diagram:} We study the dynamical phase diagram that arises after a quench of the transverse field. Concretely, the system is initialized in the ground-state of $\hat{H}_{\mathrm{D}}$ at fixed $g$ and $\delta$ with $\Omega = 0$, such that $\vert \psi(0) \rangle = \vert (-N/2)_z \rangle \otimes \vert \alpha_{s} \rangle$, and the transverse field is then quenched to a final value $\Omega \neq 0$. To garner insight into the dynamics we first study the classical model (Figs.~1-3) before probing the role of quantum fluctuations. The classical limit of the Dicke model is equivalent to solving the Heisenberg equations of motion for operators under a mean-field approximation, wherein expectation values are factorized according to $\langle\hat{\mathcal{O}}_1(t)\hat{\mathcal{{O}}}_2(t)\rangle=\langle\hat{\mathcal{O}}_1(t)\rangle \langle\hat{\mathcal{{O}}}_2(t)\rangle$ \cite{SM}. For brevity we adopt the notation $\mathcal{O} \equiv \langle \hat{\mathcal{O}}(t) \rangle$ herein.

It is convenient to study dynamics in terms of two variables: $\tilde{g} = 2g/\sqrt{\delta\Omega} \equiv g/g_{\mathrm{QPT}}$ and $\delta/\Omega$. The former characterizes the effective strength of the spin-boson interaction relative to the single-particle terms, while the latter describes the relative energy scales of spin and bosonic excitations and loosely expresses the relative importance of each degree of freedom to the dynamics. In the limit $\delta/\Omega \gg 1$ and $\tilde{g} \sim 1$ the dynamics are equivalent to a spin model described by the Lipkin-Meshkov-Glick (LMG) Hamiltonian, $\hat{H}_{\mathrm{eff}} = (\chi/N)\hat{S}^2_z + \Omega\hat{S}_x$ with the boson-mediated interaction characterized by $\chi \equiv 4g^2/\delta$ \cite{Arghavan_2017,Lang_2018}. For $\delta/\Omega \ll 1$ and $\tilde{g} \sim 1$ the spins are instead slaved to the dynamics of the dominant bosons \cite{SM}. However, in this limit a simple boson-only Hamiltonian is not generally applicable \cite{DiegoFollowUp}. Lastly, for $\delta/\Omega \sim 1$ the dynamics is complicated and involves both degrees of freedom. Herein, we refer to the integrable limits $\delta/\Omega \gg 1$ and $\delta/\Omega \ll 1$ as the spin-dominated (SDR) and boson-dominated (BDR) regimes respectively, and the non-integrable case of $\delta/\Omega \sim 1$ as the resonant regime (RR). 

\begin{figure}[tb]
    \includegraphics[width=8cm]{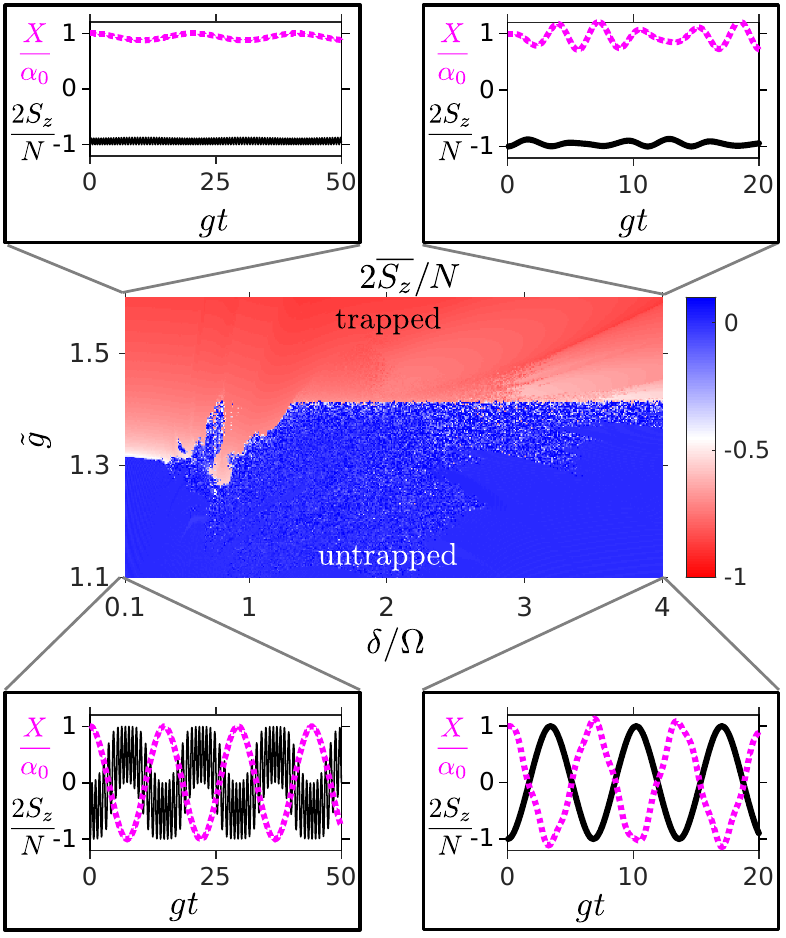}
    \caption{Dynamical phase diagram for time-averaged order parameter $\overline{S_z}$ (center) and typical time traces (surrounding) for the Dicke model. Parameters of time-traces are (clockwise from top left): $(\delta/\Omega, \tilde{g}) = (0.1,2)$, $(\delta/\Omega, \tilde{g}) = (4,2)$, $(\delta/\Omega, \tilde{g}) = (4,1)$ and $(\delta/\Omega, \tilde{g}) = (0.1,1)$. Time-averaged $\overline{S_z}$ is obtained up to a maximum time $g T = 10^4$.}
    \label{fig:Fig1}
\end{figure}

The typical dependence of the dynamics on $\tilde{g}$ is shown in the time-traces of Fig.~\ref{fig:Fig1} for SDR and BDR. In both limits we identify that the dynamics can be characterized as either \emph{trapped} or \emph{untrapped}. The former occurs when the spin-boson interaction dominates the Hamiltonian and leads to a locking of the spins and bosons close to their initial configuration. In the SDR this has been interpreted as a self-generated detuning $\propto \langle\hat{S}_z\rangle\hat{S}_z$ that locks out rotations due to the transverse field \cite{Muniz2020}. Conversely, the untrapped dynamics are characterized by large coherent oscillations in $S_z$ and $X = \frac{1}{2}\langle \hat{a} + \hat{a}^{\dagger} \rangle$ dominated by the single-particle terms of the Hamiltonian. 

We classify the dynamical phases and identify a DPT using a pair of interchangeable time-averaged order parameters $\overline{S_z} = \lim_{T\to\infty}(1/T)\int_0^T S_z(t) dt$ and $\overline{X} = \lim_{T\to\infty}(1/T)\int_0^T X(t) dt$. The trapped phase is defined by non-zero $\overline{S_z} \neq 0$ and $\overline{X} \neq 0$, while in the untrapped phase $\overline{S_z} = \overline{X} = 0$. The phases are separated by a critical coupling [main panel of Fig.~\ref{fig:Fig1}]: i) $\tilde{g}^{s}_{\mathrm{DPT}} \approx \sqrt{2}$ in the spin-dominated regime  and ii) $\tilde{g}^{b}_{\mathrm{DPT}} \approx 3^{1/4}$ in the boson-dominated regime. The dynamical phases in the LMG model have recently been observed in a cavity-QED quantum simulator \cite{Muniz2020}.

\begin{figure}[tb]
    \includegraphics[width=8cm]{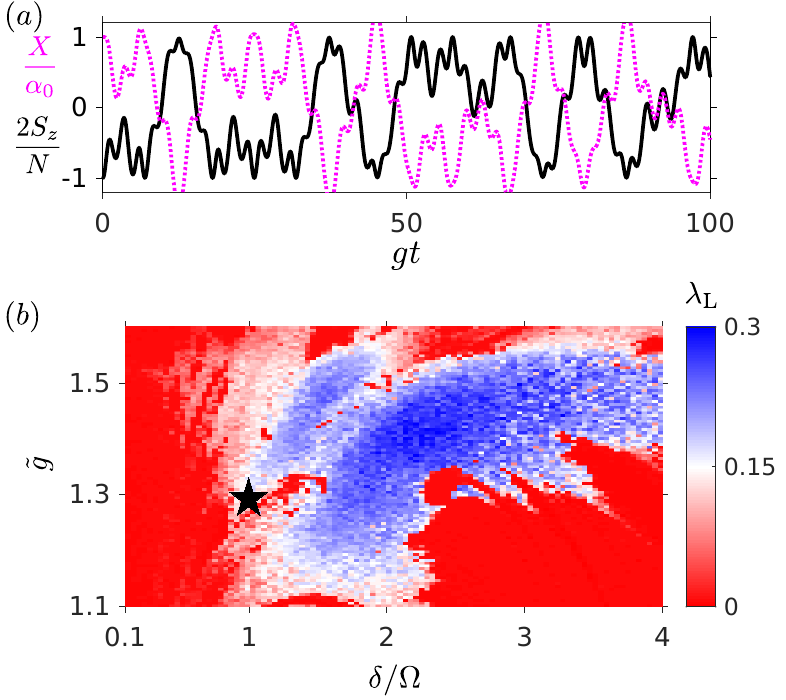}
    \caption{(a) Typical time-trace near the transition in the resonant region, $\tilde{g} \approx 1.299$ and $\delta/\Omega = 1$. (b) Characterization of phase-space via Lyapunov exponent $\lambda_{\mathrm{L}}$. Star indicates parameter regime of time-trace.}
    \label{fig:Fig2}
\end{figure}

In the non-integrable RR, $\delta/\Omega \sim 1$, the dynamical phase diagram is more complex. The main panel of Fig.~\ref{fig:Fig1} illustrates that the time-averaged order parameter becomes noisy in the untrapped phase. In this regime  typical time-traces [Fig.~\ref{fig:Fig2}(a)] feature  erratic oscillations in both spin and boson observables and there exist short periods where the system abruptly becomes re-trapped. This behaviour signals a chaotic dynamical phase \cite{lerose2018dpt,lerose2019dpt} that arises due to the known non-integrability of the Dicke model for $\tilde{g} > 1$ and $\delta \sim \Omega$ \cite{Altland_PRL_2012,RLS_scrambling_2019,Chavez_2016,Emary2003_PRL}.
 
We support this by computing the Lyapunov exponent $\lambda_{\mathrm{L}}$ \cite{RLS_scrambling_2019} in Fig.~\ref{fig:Fig2} \Rev{(see also Ref.~\cite{SM}), as a function of $\tilde{g}$ and $\delta/\Omega$ for the initial condition corresponding to $\vert\psi(0)\rangle$. Chaos, and thus non-integrability, is signalled by $\lambda_{\mathrm{L}} > 0$ \cite{strogatz2014nonlinear} in regions of parameter space that qualitatively coincide with the noisy behaviour of $\overline{S_z}$ (Fig.~\ref{fig:Fig1}).} \Rev{We also find evidence of small regions where the chaotic dynamical phases exist but $\lambda_L \to 0$ (within numerical error). This is not contradictory as the underpinning requirement for the so-called chaotic dynamical phase to exist is in fact the non-integrability of the Dicke model due to the coupling of the spin and boson degrees of freedom.}

\begin{figure}[tb]
    \includegraphics[width=8cm]{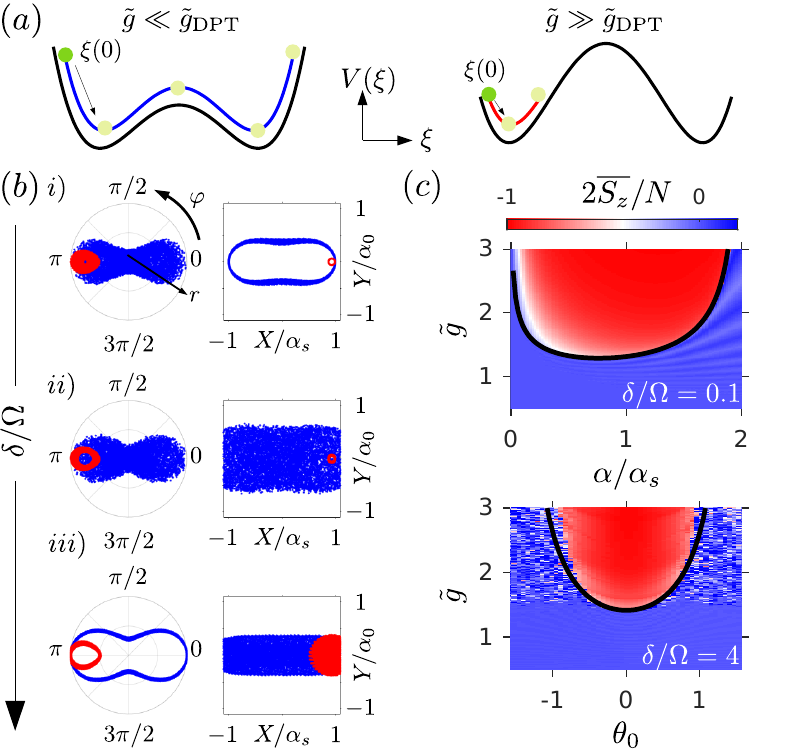}
    \caption{(a) In the integrable limits the DPT can be described as a particle (green markers) in a $1$D potential with co-ordinate $\xi = S_z$ ($\delta/\Omega \gg 1$) or $\xi = X$ ($\delta/\Omega \ll 1$). At small $\tilde{g} \ll \tilde{g}_{DPT}$ (untrapped phase, left) the particle has sufficient energy to traverse both wells of the potential, whereas for large $\tilde{g} \gg \tilde{g}_{DPT}$ (trapped phase, right) the particle remains energetically confined to a single well. (b) Typical trajectories in phase-space for trapped (red) and untrapped (blue) phases at $\delta/\Omega = (0.1, 1, 10)$ (top to bottom). For the spin phase-space we use co-ordinates $r=1 + 2S_x/N$ and $\varphi = \mathrm{arctan}(S_y/S_z)$ (c) Dynamical phase diagram as a function of initial state. At small detuning $\delta/\Omega = 0.1$ we vary the boson amplitude $\alpha$ (top panel), and at large detuning $\delta/\Omega = 4$ we vary the tipping angle $\theta_0$ relative to the south pole of the Bloch sphere ($S_z = -N/2$) (bottom panel).}
    \label{fig:Fig3}
\end{figure}

The dynamical phase diagram of SDR and BDR is captured by an effective model involving only the dominant degree of freedom \Rev{\cite{SM,sainz2008quantum,Liberti_2006,Keeling_2010,Bastarrachea_Magnani_2017,Relano_2016}}. Specifically, the mean-field dynamics are reduced to an equivalent picture of a classical particle with co-ordinate $\xi = S_z$ (spin-dominated) or $\xi = X$ (boson-dominated) confined within a $1$D potential $V(\xi)$ that depends only on $\tilde{g}$, $\delta/\Omega$ and the initial state. Near the DPT $V(\xi)$ is well approximated by a double well potential [Fig.~\ref{fig:Fig3}(a)].

The initial position and velocity of the particle, $\xi(0)$ and $\dot{\xi}(0)$, and the relative height of the central maximum, $V(0)$, characterizes the dynamics of the system \cite{lerose2019dpt}. For $\tilde{g} > \tilde{g}_{\mathrm{DPT}}$ and the particle initially located in one well, $\xi(0) \neq 0$, the particle has insufficient mechanical energy to overcome the barrier and remains confined. On the other hand, for $\tilde{g} < \tilde{g}_{\mathrm{DPT}}$ the particle has sufficient energy to pass over the barrier and traverses freely between both wells. The former scenario describes trapped dynamics, $\overline{\xi} \neq 0$, whilst the latter describes the untrapped phase, $\overline{\xi} = 0$. The critical point is defined as the condition for which the particle first surpasses the central barrier and we find $\tilde{g}^{s}_{\mathrm{DPT}} = \sqrt{2}$ and $\tilde{g}^{b}_{\mathrm{DPT}} = 3^{1/4}$, respectively, in agreement with Fig.~\ref{fig:Fig1}.


\Rev{The DPT in the SDR coincides with the well known EQPT of the Dicke Hamiltonian \cite{Perez_2011,Brandes_2013,Hirsch_2014}, i.e., the energy of the initial state matches the EQPT critical energy, $E_0\equiv \langle \psi_0 \vert \hat{H} \vert \psi_0 \rangle\vert_{\tilde{g}^{s}_{\mathrm{DPT}}} = E_c$. The EQPT features a non-analyticity in eigenstate observables at $E_c$ that can be intimately related to the behaviour of $\bar{S_z}$ at the DPT and is related to a saddle point that emerges in the classical phase-space \cite{Brandes_2013,Hirsch_2014}. In the BDR the Dicke model has been predicted to feature a set of EQPTs at energies $\{E^{(b)}_c\}$ that arise in distinct sectors of the energy spectrum labelled by values of an emergent conserved quantity, which restores the integrability of the model \cite{Bastarrachea_Magnani_2017,Relano_2016}. In the large $N$ limit our state occupies a single sector of the spectrum and at   $\tilde{g}^{b}_{\mathrm{DPT}} < \tilde{g}^{s}_{\mathrm{DPT}}$ the energy of the initial state $E_0 \equiv \langle \psi_0 \vert \hat{H} \vert \psi_0 \rangle\vert_{\tilde{g}^{b}_{\mathrm{DPT}}}$ matches a different  EQPT critical energy  $E^{(b)}_c = -\sqrt{3}\Omega N/4$, which is consistent with the observed shift in the DPT critical point.}


The potential model is further evidenced by inspecting typical trajectories of trapped and untrapped cases in the classical phase-space, as shown in Fig.~\ref{fig:Fig3}(b) for the three regimes. In BDR and SDR we observe well defined orbits in the dominant degree of freedom that are centered around fixed point(s) in phase-space corresponding to the minima of the potential well(s) \cite{SM}. The trajectories of the complementary slaved observables show similar behaviour, but tend to densely fill out the available phase-space due to fast micromotion on top of the slower orbits arising from the enslavement to the dominant degree of freedom (see also Fig.~\ref{fig:Fig1}). In the resonant regime the potential description breaks down as no single degree of freedom dominates. Consequently, we observe a lack of clear orbits in phase-space, although the erratic trajectories do fill out distinct volumes of phase-space in each dynamical phase.

DPTs in isolated systems are intrinsically dependent on initial conditions \cite{Muniz2020}. We demonstrate this by probing initial states of the form $\vert \psi(0) \rangle = \vert (-N/2)_{\mathbf{n}} \rangle \otimes \vert \alpha \rangle$ where $\alpha \in \mathbb{R}$ and we define $\hat{S}_{\mathbf{n}} \vert m_{\mathbf{n}} \rangle = m_{\mathbf{n}} \vert m_{\mathbf{n}} \rangle$ for $\hat{S}_{\mathbf{n}} = \mathbf{\hat{S}}\cdot \mathbf{n}$ with $\mathbf{n} = (0,\mathrm{sin}(\theta_0),\mathrm{cos}(\theta_0))$ defined by the tipping angle $\theta_0$ from the south pole of the collective Bloch sphere. In Fig.~\ref{fig:Fig3}(c) we probe the dependence of $\tilde{g}_{\mathrm{DPT}}$ on the initial amplitude $\alpha$ in the BDR, while in the SDR we vary the initial tipping angle $\theta_0$. We observe good agreement with analytic predictions for $\tilde{g}_{\mathrm{DPT}}$ based on the potential model \cite{SM}. Small deviations are observed for $\theta \approx \pm \pi/2$ and $\alpha/\alpha_{s} \to 0$, which correspond to initializing a particle near the central maximum of the potential such that even a small off-resonant exchange of energy with the complementary degree of freedom is enough for the particle to overcome the barrier.

\begin{figure}[tb]
    \includegraphics[width=8cm]{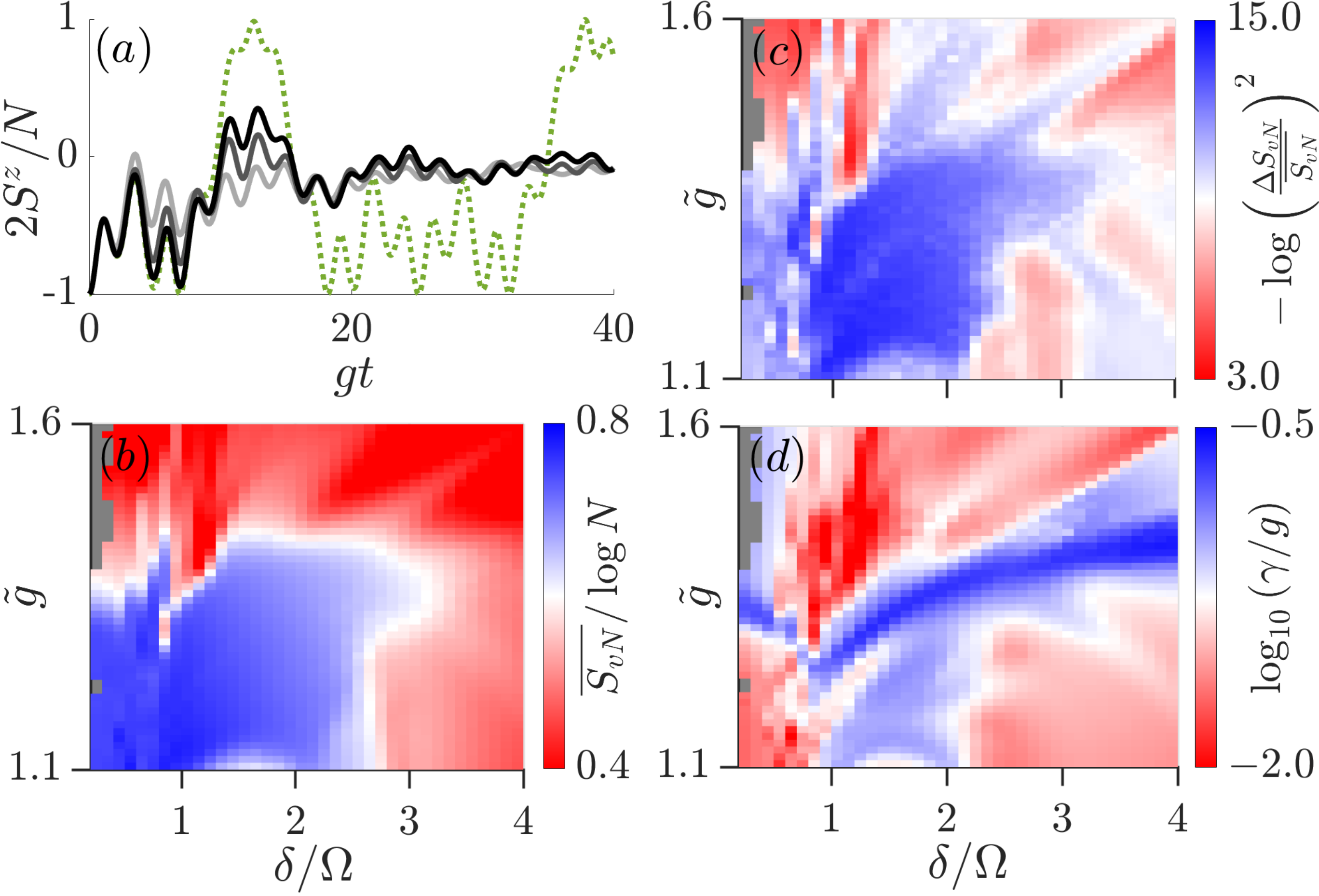}
    \caption{Signatures of DPT in quantum dynamics (see Ref.~\cite{SM} for details). (a) Time traces of quantum dynamics in the resonant regime, $\tilde{g} = 1.3$ and $\delta/\Omega = 1$, for $N = 40$ (light gray), $200$ (gray), and $600$ (black). Mean-field solution (green) is shown for reference. (b) Time-averaged entanglement entropy $\overline{ S_{\mathrm{vN}} } = (1/T)\int_{t_0}^{T+t_0} ~dt S_{\mathrm{vN}}(t)$ for $S_{\mathrm{vN}} = -\mathrm{Tr}[\hat{\rho}_s \mathrm{log}(\hat{\rho}_s)]$ where $\hat{\rho}_s$ is the reduced density matrix of the spins. 
    (c) Temporal fluctuations of entanglement entropy, $(\Delta S_{\mathrm{vN}})^2 = (1/T)\int_{t_0}^{T+t_0} ~dt [S_{\mathrm{vN}}(t) - \overline{ S_{\mathrm{vN}} }]^2$, normalized by $\overline{S_{\mathrm{vN}}}$. (d) Growth rate $\gamma$ of quantum fluctuations, obtained by an empirical fit to $\langle (\Delta\hat{S}_z)^2 \rangle \propto (1 - e^{-\gamma t})$. Data in (b)-(d) is obtained for $N=600$. Time averages in (b) and (c) are computed over a window $t\in[t_0,T+t_0]$ for $gt_0, gT = 150$, so as to reduce transient effects in steady state estimates. Gray regions correspond to excluded data.}
    \label{fig:Fig4}
\end{figure}

\noindent{\it Quantum dynamics:} Using an efficient exact diagonalization method \cite{SM} we are able to simulate the quantum dynamics and observe mean-field signatures of the DPT that survive quantum noise. For example, in Fig.~\ref{fig:Fig4}(a) we probe a typical time-trace of $\langle \hat{S}_z \rangle$ in the resonant regime [same as Fig.~\ref{fig:Fig2}(a)] for experimentally realistic $N=40,200,600$ and find that it is possible to gradually observe the signature re-trapping dynamics with increasing $N$, before quantum fluctuations dephase $\langle \hat{S}_z \rangle \to 0$.

Conversely, pure quantum measures such as entanglement and quantum correlations also show clear signatures of the DPT, chaos and non-integrability. Figure \ref{fig:Fig4}(b) shows that the critical point of the DPT is signaled in the pronounced build-up of spin-boson entanglement in the untrapped phase, quantified by the entanglement entropy $S_{\mathrm{vN}} = -\mathrm{Tr}[\hat{\rho}_s \mathrm{log}(\hat{\rho}_s)]$ where $\hat{\rho}_s$ is the reduced density matrix of the spins. For $\delta/\Omega \lesssim 1$ the time-averaged entropy $\overline{ S_{\mathrm{vN}} } = (1/T)\int_{t_0}^{T+t_0} ~dt S_{\mathrm{vN}}(t)$ demarcates the trapped and untrapped phases consistent with the order parameter $\overline{S_z}$, with  $\overline{ S_{\mathrm{vN}} }_{\mathrm{trapped}} \ll \overline{ S_{\mathrm{vN}} }_{\mathrm{untrapped}}$. The entanglement  decreases with increasing $\delta/\Omega$ as the bosons become eliminated, although we expect that, e.g., bipartite entanglement between the spins will retain indications of the DPT. Panel (c) also demonstrates that the temporal fluctuations of the entanglement vary with the underlying integrability of the Hamiltonian \cite{Kaplan_2020,Kiendl_2017}. Comparing to the Lyapunov exponent in Fig.~\ref{fig:Fig2}, we see a correlation between regions of chaotic (non-integrable) dynamics and suppressed temporal fluctuations of $S_{\mathrm{vN}}(t)$ (relative to $\overline{ S_{\mathrm{vN}} }$).

Panel (d) evidences that the critical region of the DPT can be diagnosed by the rapid growth of quantum fluctuations. By empirically fitting $\langle (\Delta\hat{S}_z)^2 \rangle = \langle \hat{S}_z^2 \rangle - \langle \hat{S}_z \rangle^2 \propto (1 - e^{-\gamma t})$ we find that the rate at which quantum fluctuations buildup, characterized by $\gamma$, is largest for $\tilde{g} \sim \tilde{g}_{\mathrm{DPT}}$. This is consistent with the effective potential description, as near the critical coupling the mean-field trajectory spends an increasing amount of time probing the central barrier (e.g., unstable fixed point) \cite{SM,lerose2019dpt,Lerose_2020_semiclassical,Lerose_2020}, motivating the expectation that quantum fluctuations will grow exponentially. Nevertheless, we also note that a similarly rapid growth of fluctuations is observed away from the DPT for $1 \lesssim \delta/\Omega \lesssim 2$ as a consequence of classical chaos \cite{RLS_scrambling_2019}, indicating that it is important to carefully differentiate the effects of chaos from the DPT in quantum dynamics.

\noindent{\it Experimental realization:} The full dynamical phase diagram of the Dicke model can be studied in a range of current state-of-the-art AMO quantum simulators, but most readily in arrays of trapped ions. In particular, the Dicke model was recently realized in a $2$D Penning trap configuration \cite{Bollinger_2018,Cohn_2018} where a spin-$1/2$ is encoded in two internal states of each ion and the collective center-of-mass motional mode of the ion crystal realizes the bosonic degree of freedom. A pair of lasers generates an optical dipole force that couples the internal state of each ion to the motional degree of freedom \cite{Britton2012}. A transverse field is generated by applying microwaves with tunable strength $\Omega$ that coherently drive the two internal levels and the spin-boson coupling $g$ can be in principle controlled via the applied laser power. Moreover, $\delta$ can be varied by controlling the detuning of the lasers relative to the frequency of the targeted center-of-mass mode \cite{Britton2012}. Considering the simulator reported in Refs.~\cite{Bollinger_2018,Cohn_2018}, we predict it should be possible to probe regimes $0.1 \lesssim \delta/\Omega \lesssim 10$ for $\tilde{g}\sim 1$ on timescales that are fast compared to relevant sources of single-particle decoherence of the spins \cite{SM,Bohnet2016,Arghavan_2017}. In contrast to related studies of the Dicke model in optical cavities, the trapped ion simulator has little damping of the bosonic mode and thus to an excellent approximation can be neglected. Further, ion crystals of $N\approx 200$ \cite{Bohnet2016} are possible in the $2$D geometry, which is sufficient to access the signatures of the DPT as in Figs.~\ref{fig:Fig1} and \ref{fig:Fig4} \cite{SM}. Looking ahead, implementing the Dicke model in $3$D ion crystals \cite{Mortensen2006} would open a path to even larger system sizes that are beyond the capability of numerical methods.

\noindent{\it Conclusion:} We have studied a DPT in the Dicke model and unique features arising from non-integrable and chaotic regimes of parameter space. Our numerical study indicates signatures of the DPT survive in the quantum dynamics and are accessible in current state-of-the-art AMO quantum simulators based on, e.g., trapped-ion arrays. This can motivate future investigations connecting to quantum phenomena such as, e.g., information scrambling \cite{RLS_scrambling_2019,Joshi_2020}. Additionally, studying entanglement dynamics associated with DPTs in collective systems will open new directions for the generation of metrologically useful states for quantum-enhanced sensors \cite{Gilmore2017} or frequency and time standards \cite{Brewer_2019}.

\begin{acknowledgments}
\noindent{\textit{Acknowledgements:}} We acknowledge helpful discussions with Jamir Marino and Matthew Affolter regarding the manuscript. After initial submission of our work for peer review, Armando Rela\~{n}o and Miguel A. Bastarrachea Magnani pointed out the correspondence of the DPT with emergent EQPTs in the boson-dominated regime and we are thankful for their contribution. This work is supported by the AFOSR grant FA9550-18-1-0319, by the DARPA and ARO grant W911NF-16-1-0576, the ARO single investigator award W911NF-19-1-0210, the NSF PHY1820885, NSF JILA-PFC PHY-1734006 and NSF QLCI-2016244 grants, by 
the DOE, Office of Science, National Quantum Information Science Research Centers, Quantum Systems Accelerator (QSA) and by NIST.
\end{acknowledgments}

\bibliography{library}

\newpage 

\onecolumngrid
\vspace{\columnsep}
\begin{center}
\textbf{\large Supplemental Material: Characterizing the dynamical phase diagram of the Dicke model via classical and quantum probes}
\end{center}
\vspace{\columnsep}

\setcounter{equation}{0}
\setcounter{figure}{0}
\setcounter{table}{0}
\setcounter{page}{1}
\makeatletter
\renewcommand{\theequation}{S\arabic{equation}}
\renewcommand{\thefigure}{S\arabic{figure}}

\section{Classical limit of Dicke model: Mean-field equations of motion}
The mean field equations of motion (EOM) can be derived from the Heisenberg equations of motion by replacing operators $\hat{a}$ and ($\hat{S}_x$, $\hat{S}_y$, $\hat{S}_z$) with the classical variables $\beta/\sqrt{N}$ and ($s_x$, $s_y$, $s_z$)/($N/2$), respectively:
\begin{align}
    \begin{split}
\dot{\beta}&=-i\delta \beta -ig s_z\\
\dot{s}_x&=-2g(\beta+\beta^*)s_y\\
\dot{s}_y&=2g(\beta+\beta^*)s_x-\Omega s_z\\
\dot{s}_z&=\Omega s_y.
\end{split}\end{align}
These should be combined with the initial conditions $\beta(0)=g\beta_0/\delta$ and $[s_x(0),s_y(0),s_z(0)]=[\sin\theta_0,0,-\cos\theta_0]$, where $\beta_0$ is a quantity of order $1$ and $\theta_0$ is the initial tipping angle of the spins with respect to the $-z$ direction in the $xz$ plane. Since the initial condition for $\beta$ is parameter dependent, we define $\tilde{\beta}=\delta\beta/g$ and $\tilde\beta(0)=\beta_0$ so that the equations of motion take a more insightful form:  
\begin{align}\label{eqn:SMMeanFieldEOM}
    \begin{split}
\dot{\tilde\beta}&=-i\delta(\tilde{\beta} + s_z)\\
\dot{s}_x&=-\Omega \bigg[\frac{\tilde{g}^2}{2}(\tilde{\beta}+\tilde{\beta}^*)s_y\bigg]\\
\dot{s}_y&=\Omega\bigg[\frac{\tilde{g}^2}{2}(\tilde{\beta}+\tilde{\beta}^*)s_x-s_z\bigg]\\
\dot{s}_z&=\Omega s_y,
\end{split}\end{align}
where $\tilde{g}=g/g_{QPT}=2g/\sqrt{\delta\Omega}$ is a normalized coupling constant. This form of the EOM allows us to unambiguously analyze the spin- ($\delta/\Omega) \to \infty$) and boson-dominated ($\delta/\Omega \to 0$) limits at fixed $\tilde{g}$.

\section{Analytic treatment of DPT in integrable regimes \label{sec:AnalyticDPT}}
In this section we present a detailed outline of the classical particle in a potential description of the DPT in the integrable spin- and boson-dominated limits. We derive each case separately, drawing from the above mean-field EOM [Eq.~(\ref{eqn:SMMeanFieldEOM})], and present the dependence of the critical point of the DPT on the initial state. 

\subsection{Spin-dominated regime: $\delta/\Omega \gg 1$}
First, we address the limit where the spins are the dominant degree of freedom. To derive an effective model of the dynamics we take the limit $\delta/\Omega\to \infty$ but keeping $\Omega$ fixed and formally eliminate the bosonic degree of freedom so we are left with a simpler pure spin description. This is achieved by observing from Eq.~(\ref{eqn:SMMeanFieldEOM}) that the bosons will evolve on a much faster timescale ($1/\delta$) than the spins ($1/\Omega$). Thus, the dynamics of the bosons is composed of two different contributions: a slow drift created by the slow time evolution of $s_z$ plus very fast oscillations about it. The slow dynamics is characterized by the locally averaged value $\tilde\beta$, which is obtained by setting $\dot{\tilde\beta}=0$ in Eq.~(\ref{eqn:SMMeanFieldEOM}): 
\begin{equation}
    \tilde\beta_{\text{av}}=-s_z.
\end{equation}

As a first approximation, the spin degree of freedom will only be sensitive to the dynamics of $\tilde\beta_{\text{av}}$, such that the spins evolve under a closed set of equations of motion (EOM):
\begin{align}\label{eqn:SMMeanFieldLMG}
\begin{split}
\dot{s}_x&=\Omega \tilde{g}^2s_z s_y\\
\dot{s}_y&=-\Omega\big(\tilde{g}^2s_z s_x+s_z\big)\\
\dot{s}_z&=\Omega s_y.
\end{split}\end{align}
These equations can equivalently be derived from a classical Hamiltonian
\begin{eqnarray}
    h=\Omega\bigg(-\frac{\tilde{g}^2}{2}s_z^2+s_x\bigg),
\end{eqnarray}
with the standard angular momentum Poisson brackets: $\{s_i,s_j\}=\epsilon_{ijk}s_k$. 

The energy $h$ is a conserved quantity defined with respect to the initial condition of the spins, $[s_x(0),s_y(0),s_z(0)]=[\sin\theta_0,0,-\cos\theta_0]$:
\begin{equation}
    h_0=\Omega\bigg(-\frac{\tilde{g}^2}{2}\cos\theta_0^2+\sin\theta_0\bigg).
\end{equation}
Similarly, the total spin length, $s_x^2+s_y^2+s_z^2=1$, is also conserved under the dynamics. Together, this allows us to eliminate $s_x$ (using conservation of $h$) and $s_y$ (using the EOM), to obtain a single equation for $s_z$:
\begin{align}\begin{split}
    \frac{(\dot{s_z}^2)}{2}+ \underbrace{\frac{\Omega^2}{2}\bigg\{\bigg[\frac{\tilde{g}^2}{2}(s_z^2-\cos\theta_0^2)+\sin\theta_0\bigg]^2+s_z^2-1\bigg\}}_{V(s_z)}&=0.
\end{split}\end{align}
This differential equation describes the motion of a classical particle in a one-dimensional potential $V(s_z)$. The dynamics of the particle is dictated by the form of the potential and the initial condition: The particle is initialized at $s_z(0)=-\cos\theta_0$ with potential energy $V(s_z(0)) = 0$ and kinetic energy $\dot{s_z}(0)^2/2 = 0$. For $\tilde{g} \to 0$, the particle can freely move from $s_z=-\cos\theta_0$ to $s_z=\cos\theta_0$ and back. This describes the untrapped regime. As $\tilde{g}$ is increased, $V(s_z)$ develops a maximum at $s_z=0$. The height of this central barrier increases until it matches the initial mechanical energy of the particle for a critical coupling 
\begin{equation}
    \tilde{g}^s_{\text{DPT}}(\theta_0)=\sqrt{2}\sqrt{\frac{1+\sin\theta_0}{\cos\theta_0^2}} . 
\end{equation}
\begin{figure}
    \centering
    \includegraphics[width=0.8\textwidth]{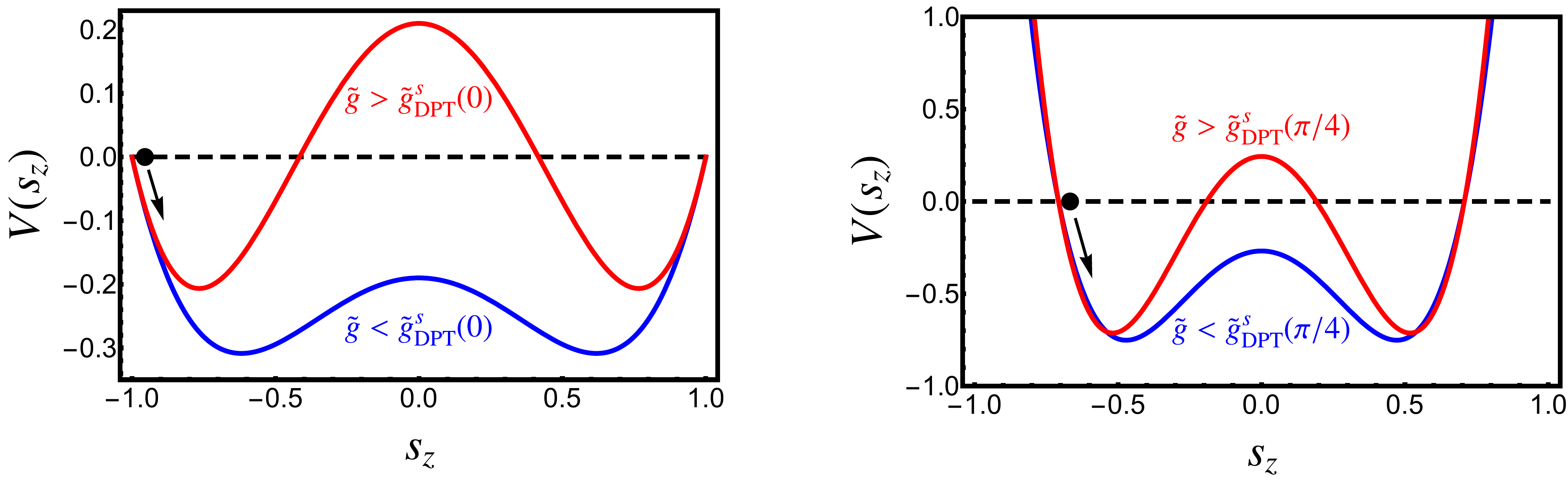}
    \caption{Effective potential for the spin degree of freedom when $\delta\gg \Omega$ and $\theta_0=0$ (left) or $\theta_0=\pi/4$ (right). The ``particle" begins at $s_z=-\cos(\theta_0)$ and then moves to the right. When $\tilde{g}<\tilde{g}^s_{\text{DPT}}(0)$ (blue) the central barrier is below the initial ``energy" and so $s_z$ oscillates between $-\cos(\theta_0)$ and $\cos(\theta_0)$. When $\tilde{g}>\tilde{g}^s_{\text{DPT}}$ (blue) the particle cannot overcome the central barrier and so $s_z<0$ for all times. As illustrated by the two contrasting panels, the exact form of the potential depends on the initial state of the system (parameterized by $\theta_0$ in this instance), and in the spin-dominated regime this solely introduces the dependence of the critical coupling $\tilde{g}^s_{\text{DPT}}(\theta_0)$ on the initial state.}
    \label{fig:supp_effpotential}
\end{figure}
For $\tilde{g} \geq \tilde{g}^s_{\mathrm{DPT}}(\theta_0)$ the particle is confined to $s_z\leq0$ for all times (see Fig.~\ref{fig:supp_effpotential}). This is the trapped regime. In the case $\theta_0=0$, which is the primary focus of the main text, $\tilde{g}^s_{\text{DPT}}=\sqrt{2}$.

\subsection{Boson-dominated regime: $\delta/\Omega \ll 1$}
Here, we address the limit where the bosons are the dominant degree of freedom. We direct the interested reader to Refs.~\cite{Liberti_2006,sainz2008quantum,Bastarrachea_Magnani_2017,Relano_2016} for related approaches in the literature. To derive an effective model of the dynamics we take the limit $\delta/\Omega\to 0$ keeping $\delta$ fixed. Complementary to the prior analysis, the spin degree of freedom now has a very short evolution timescale ($1/\Omega$) as compared to the bosons ($1/\delta$). Hence, from the perspective of the spins, the $\tilde\beta$ term in their equations of motion [see Eq.~(\ref{eqn:SMMeanFieldEOM})] is static. Hence, the spin dynamics can be understood as a fast precession with respect to a ``static" $\tilde\beta$-dependent axis. Furthermore, the slow time evolution of $\tilde\beta$ induces adiabatic evolution of the spins on top of their fast precession, with instantaneous axis of rotation given by
\begin{eqnarray}
    \mathbf{n}_t\equiv \frac{1}{\sqrt{1+\tilde{g}^4\tilde{\beta}_R^2}}\big(1,0,\tilde{g}^2\tilde{\beta}_R\big),
\end{eqnarray}
where $\tilde{\beta}_R=(\tilde{\beta}+\tilde{\beta}^*)/2$ is the real part of $\tilde{\beta}$ and $\mathbf{n}_t$ owes its time dependence to $\tilde{\beta}_R$.

The drastic difference in time-scales between the spin and bosonic degrees of freedom means that the bosons are only sensitive to the local average of the spin vector (e.g., we ignore the rapid oscillations on $1/\Omega$ timescales) $\mathbf{s}$:
\begin{eqnarray}
    \mathbf{s}_{\text{av}}= (\mathbf{n}_{t=0}\cdot \mathbf{s}_0)\,\mathbf{n}_t,
\end{eqnarray}
with $\mathbf{s}_0 \equiv \mathbf{s}(0)$ the initial condition. For simplicity of analysis in the following, we will take $\theta_0 = 0$ (relevant for the main text). Then, the boson equation of motion becomes
\begin{equation}
    \dot{\tilde{\beta}}=-i\delta \Bigg(\tilde\beta-\frac{\tilde{g}^2\beta_0}{\sqrt{1+\tilde{g}^4\beta_0^2}}\frac{\tilde{g}^2\tilde{\beta}_R}{\sqrt{1+\tilde{g}^4\tilde{\beta}_R^2}}\Bigg),
\end{equation}
which can be recast in terms of an effective potential by eliminating the imaginary part of $\tilde{\beta}$:
\begin{equation}
    \frac{\dot{\tilde{\beta}}_R^2}{2}+\underbrace{\delta^2\Bigg(\frac{\tilde{\beta}_R^2}{2}-\frac{\beta_0\sqrt{1+\tilde{g}^4\tilde{\beta}_R^2}}{\sqrt{1+\tilde{g}^4\beta_0^2}}+\beta_0-\frac{\beta_0^2}{2}\bigg)}_{V(\tilde{\beta}_R)}=0.
\end{equation}
The analysis of this differential equation follows identically to that of the spin-dominated regime. For $\tilde{g} \to 0$ the particle moves freely between $\pm\beta_0$, which corresponds to the untrapped regime. As $\tilde{g}$ increases, a maximum develops at $\tilde{\beta}_R=0$ and eventually acts as a barrier that keeps $\beta_0>0$ for all times. This occurs at the critical coupling
\begin{equation}
    \tilde{g}^b_{\text{DPT}}(\beta_0)=\left[\frac{4-\beta_0}{\beta_0(2-\beta_0)^2}\right]^{1/4}. \label{eqn:gb_DPT}
\end{equation}
For $\beta_0=1$, we recover $\tilde{g}_{\text{DPT}}=3^{1/4}$ as per the result quoted in the main text. An important difference with the spin potential is that in this case the form of the effective boson potential depends on the initial state of the spins as well as the initial state of the bosons, whereas the spin potential is insensitive to the initial state of the bosons. While our result here, Eq.~(\ref{eqn:gb_DPT}), specializes to $\theta_0=0$, we emphasize that the calculation is easily generalized for arbitrary $\theta_0$.

\section{Comparison to EQPT}
The Dicke model has been intensely studied with regards to the presence of a transition in the excited-state spectrum of the Hamiltonian (EQPT). Not only was the nature of the EQPT subject to initial debate \cite{Perez_2011,Brandes_2013,Puebla_2014}, although it is now accepted as a true quantum phase transition, but also there was initially speculation as to whether the EQPT might be connected to various non-equilibrium features of the model such as the emergence of chaos in the energy spectrum \cite{Perez_2011,Chavez_2016}. 

In general, the Dicke model features an EQPT for $\tilde{g} > 1$ (e.g., when the coupling is large enough that the ground-state corresponds to the superradiant phase) at a critical energy $E_c = -\Omega N/2$. It is signaled by a non-analyticity in the density of states, in particular a logarithmic divergence in the first derivative \cite{Brandes_2013}, as well as singularities in $\langle \hat{S}_x \rangle$ and $\langle \hat{a}^{\dagger} \hat{a} \rangle$ computed with respect to the energy eigenstates \cite{Perez_2011}. The EQPT can be explicitly connected to the classical model in the large-$N$ limit by noting that it arises due to the emergence of a saddle point in the classical phase-space \cite{Brandes_2013,Hirsch_2014}.

The DPT we study is quantitatively connected to this EQPT in the spin-dominated regime: We find that the energy of the initial state, $E_0 = \langle \psi(0) \vert \hat{H}_{\mathrm{D}} \vert \psi(0)\rangle = -(\tilde{g}^2/2)\Omega N/2$, coincides with the critical energy $E_c = -\Omega N/2$ at $\tilde{g}^{s}_{\mathrm{DPT}} = \sqrt{2}$. Indeed, going a step further, it is straightforward to prove that the DPT in the spin-dominated regime always precisely occurs when $E_0 = E_c$ for any initial state (e.g, arbitrary $\theta_0$, as covered in the previous treatment of the DPT by the particle in a potential model). This close correspondence is consistent with prior work in the literature \cite{Puebla_2014}, which suggested that EQPTs might be accessible through a type of DPT driven by the breaking of the parity symmetry in the energy spectrum.

In the boson-dominated regime the situation is made complex by the fact the model becomes approximately integrable. Specifically, previous work \cite{Bastarrachea_Magnani_2017,Relano_2016} has demonstrated that the Dicke model in the limit $\delta/\Omega \to 0$ features an emergent conserved quantity, 
\begin{equation}
    \hat{S}^{\prime}_x = \frac{\hat{S}_x + \tilde{g}\sqrt{\frac{\delta}{\Omega}}\frac{\hat{a}+\hat{a}^{\dagger}}{\sqrt{N}}\hat{S}_z}{\sqrt{1 + \tilde{g}^2\frac{\delta}{\Omega}\frac{\left( \hat{a}+\hat{a}^{\dagger} \right)^2}{N}}},
\end{equation}
which splits the energy spectrum into distinct sectors. Within each sector, labeled by the value $m_x^{\prime}$ of the conserved quantity, the model features an EQPT at the critical energy $E^{b}_c(m_x^{\prime}) = -\Omega m^{\prime}_x$. In the limit of large $N$, our initial state corresponds to the sector with
\begin{equation}
    m^{\prime}_x = \frac{\tilde{g}^2}{\sqrt{1 + \tilde{g}^4}} \frac{N}{2} ,
\end{equation}
and we find that the energy of our initial state at the BDR DPT then precisely coincides with the EQPT, $E_0\vert_{\tilde{g}^{b}_{\mathrm{DPT}}} = E^{b}_c(m_x^{\prime})\vert_{\tilde{g}^{b}_{\mathrm{DPT}}} = -\sqrt{3}\Omega N/4$.

\section{Relation between Lyapunov exponent and chaotic dynamical phase}
\begin{figure}
    \centering
    \includegraphics[width=0.45\textwidth]{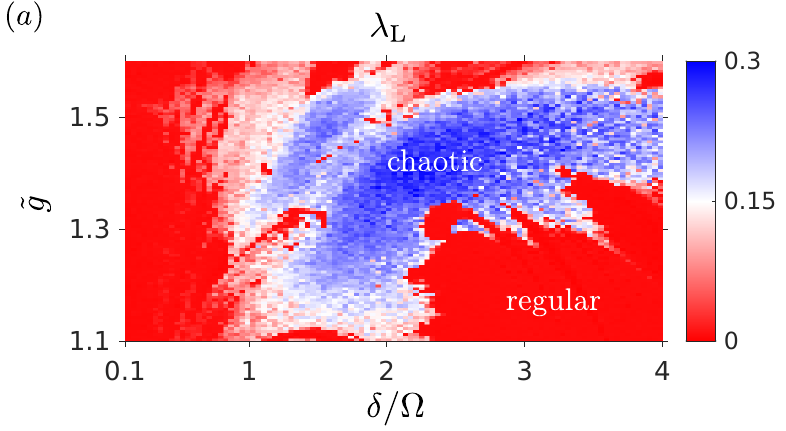}
    \includegraphics[width=0.45\textwidth]{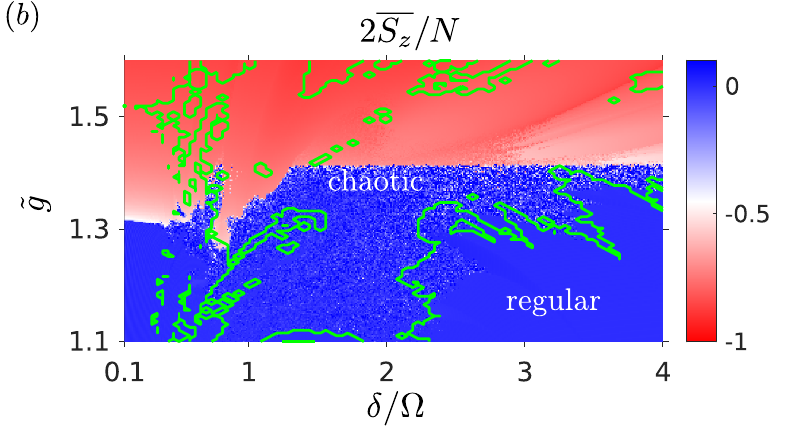}
    \caption{(a) Lyapunov exponent $\lambda_L$, reproduced from Fig.~2 of the manuscript. (b) Classical dynamical phase diagram of Dicke model in terms of the order parameter $2\overline{S_z}/N$ (see Fig.~1 of the manuscript). The overlaid green lines indicate the contours of the Lyapunov exponent colormap at $\lambda_L = 0.01$. Crudely, indicated regions interior of the green lines indicate the region with $\lambda_L > 0$, e.g., the chaotic regime.}
    \label{fig:LambdaBoundary}
\end{figure}

In the main text we highlight that the chaotic dynamical phase can be correlated closely with the known chaos of the Dicke model, as quantified by the Lyapunov exponent (see Figs.~1 and 4, respectively). Figures \ref{fig:LambdaBoundary}(a) and (b) quantify this relation more closely. Particularly, in panel (b) we overlay the dynamical phase diagram with a contour line corresponding to $\lambda_L = 0.01$ in (a), approximately indicating that the interior region of the greend boundary corresponds to chaotic dynamics. We observe that this reasonably aligns with the noisy (e.g., speckled) patches of the dynamical phase diagram. 

We briefly note that disagreement between the region of non-zero Lyapunov exponent and the order parameter noise, such as, e.g., for $(\delta/\Omega,\tilde{g}) \approx (1.25,2.5)$ arises due to the fact that chaos is not a one-to-one indicator of non-integrability. In turn, so-called `chaotic' dynamical phases only require the underlying model to be \emph{non-integrable} \cite{lerose2018dpt,lerose2019dpt}, and so the lack of chaos for a given set of parameters does not exclude the existence of `chaotic' dynamical phases. 

\section{Numerical simulation of quantum dynamics}
We efficiently simulate the quantum dynamics of the Dicke model by solving the time-dependent Schr{\"o}dinger equation using a Krylov-subspace projection method \cite{Sidje_1998}. Using the symmetries of the system, we are able to make our numerical calculation tractable by: i) representing the spin degree of freedom in the fully symmetric subspace of Dicke states $\vert S, m_z \rangle$ with $S = N/2$, defined such that $\hat{S}^2\vert S, m_z \rangle = S(S+1)\vert S, m_z \rangle$ for $\hat{S}^2 = \sum_{\alpha = x,y,z} \hat{S}_{\alpha}^2$ and $\hat{S}_z\vert S, m_z \rangle = m_z \vert S, m_z \rangle$, and ii) representing the boson degree of freedom in a truncated Fock basis $\vert n \rangle$ where $n\in [0,n_{\mathrm{max}}]$ for some maximum occupation $n_{\textrm{max}}$. While it is straightforward to simulate the evolution using a fixed $n_{\textrm{max}}$, this may require extensive trial runs and benchmarking to ensure convergence of all relevant observables. 

To avoid this issue, and to efficiently simulate large system sizes to long times over a wide range of parameter space, we utilize a bosonic Hilbert space that dynamically grows or shrinks throughout the evolution. Initially, we set $n_{\textrm{max}}$ by bounding the error in the initial wavefunction normalization  $1 - \vert \langle \psi(0) \vert \psi(0) \rangle\vert^2 \leq \epsilon = 10^{-6}$; by checking for convergence as $\epsilon$ is varied, we have verified that this bound is sufficient to generate relative errors in the spin-boson entanglement entropy dynamics on the order of $10^{-5}$ at long times, and even smaller errors for relevant observables. Throughout the evolution, we periodically check (on a time interval $\tau$) the normalization of our wavefunction when projected onto states with boson occupations of $n_{\textrm{max}}$ and $n_{\textrm{max}}-1$. If the error introduced by the projection onto either of these sets of states exceeds $\epsilon$ at any point in this interval, we increase $n_{\textrm{max}}$ by a fixed amount $\Delta n$ and recompute the evolution over this interval, repeating until the error tolerance is satisfied. In general, we find that the short-time dynamics requires a much larger value of $n_{\textrm{max}}$ than the initial state; in contrast, however, the late-time dynamics typically require a much smaller value of $n_{\textrm{max}}$. To speed up the computation of the late-time dynamics, we also allow for a decrease in the value of $n_{\textrm{max}}$, likewise checking the projection onto all boson occupations between $n_{\textrm{max}} - \Delta n$ and $n_{\textrm{max}}$. If the projection onto this set of states is less than $\epsilon$ at any point over some interval $\tau$, we reduce $n_{\textrm{max}}$ by $\Delta n$, and truncate away the corresponding states from our Hilbert space.

\begin{figure}[tb]
    \includegraphics[width=18cm]{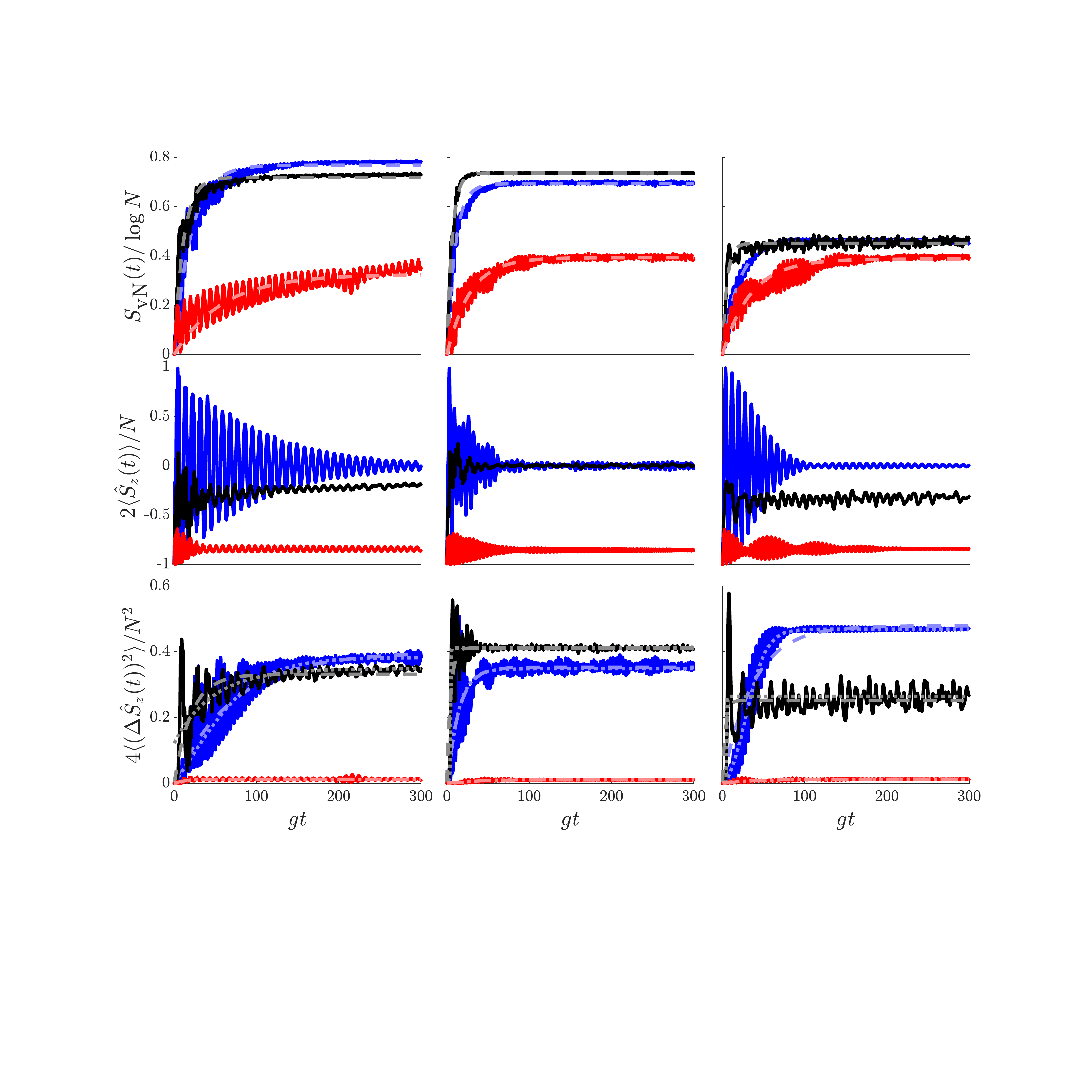}
    \caption{Typical time-traces of the entanglement entropy $S_{\textrm{vN}}$ (top row), magnetization $\langle\hat{S}^z\rangle$ (middle row), and variance $\langle (\Delta \hat{S}_z)^2\rangle$ (bottom row) for the quantum Dicke model with $N=600$, up to time $gt = 300$. Columns correspond to different values of $\delta/\Omega$, with $\delta/\Omega = 0.5$ (left column), $1.2$ (middle column), and $4.0$ (bottom row) corresponding to the boson-dominated, resonant, and spin-dominated regimes, respectively. Blue curves are for $\tilde{g} = 1.1$, corresponding to the untrapped phase, while red curves are for $\tilde{g} = 1.6$, corresponding to the trapped phase. Black curves are selected for $\tilde{g}$ occurring at the DPT for the corresponding values of $\delta/\Omega$: for $\delta/\Omega = 0.5$, $1.2$, and $4.0$, we have $\tilde{g} = 1.32$, $1.29$, and $1.43$, respectively. The dashed lines plotted alongside the entropy and variance are the corresponding fits to the function $A(1-e^{-\gamma t})$ (used to extract the timescale $\gamma$ for the variance in Fig.~4d of the main text), while the dotted lines plotted alongside the variance are the corresponding fits to the translated logistic function (see text). }
    \label{fig:SM_numerics}
\end{figure}

For all simulations in Fig.~4(b)-(d), we use $N=600$ and evolve to a fixed time $gt = 300$, initializing the system in the state $\ket{(-N/2)_z}\otimes\ket{\alpha_s}$ with $\alpha_s = (g/\delta)\sqrt{N}$ as described in the main text. In Fig.~\ref{fig:SM_numerics}, we show typical dynamics of the entanglement entropy, $S_{\mathrm{vN}} = -\mathrm{Tr}[\hat{\rho}_s \mathrm{log}(\hat{\rho}_s)]$ where $\hat{\rho}_s$ is the reduced density matrix of the spins, magnetization $\langle \hat{S}^z\rangle$, and variance $\langle(\Delta \hat{S}^z)^2\rangle = \langle\hat{S}_z^2 \rangle - \langle \hat{S}_z \rangle^2$ for various parameter regimes. To produce Fig.~4 of the main manuscript, we simulate the Dicke model for a range of $51$ values of $\tilde{g}$ and $39$ values of $\delta/\Omega$, equally-spaced between the bounds shown on the plot. For a small set of parameters, typically with $\delta/\Omega \lesssim 0.3$ and large $\tilde{g} \gtrsim 1.4$ (corresponding to large initial boson occupation), the required time and memory resources for simulation are much greater than other regions of parameter space. Thus, they are thus excluded from our results and indicated by the gray regions in Fig.~\ref{fig:SM_numerics}.

For the entanglement entropy and spin variance in Fig.~\ref{fig:SM_numerics}, we also display fits to an empirical function $A(1-e^{-\gamma t})$ with free fitting parameters $A$ (steady state value) and $\gamma$, shown as dashed lines in Fig.~\ref{fig:SM_numerics}. We generally find that the entanglement entropy $S_{vN}(t)$ is consistent with such a functional form, with the exception of transient oscillations on fast time-scales. For some parameters we find that the early time dynamics of the spin variance $\langle(\Delta \hat{S}^z)^2\rangle$ is captured more accurately by a translated logistic function, $a/(1 + e^{-c (t - t_0)}) - b$, than an exponential function; here, $a$ provides the overall scale of the function to be consistent with the observed steady state, $c$ is the logistic growth rate, $t_0$ is the function midpoint, and $b$ is an offset to allow for the fact that $\langle(\Delta\hat{S}^z)^2\rangle = 0$ at $t=0$. Fitting to this function provides an alternative estimate for the relaxation timescale of the dynamics via $\gamma_{\mathrm{logistic}} = c + 1/t_0$, where we account for both the logistic growth timescale as well as the delay in the onset of this growth.

\begin{figure}[tb]
    \includegraphics[width=18cm]{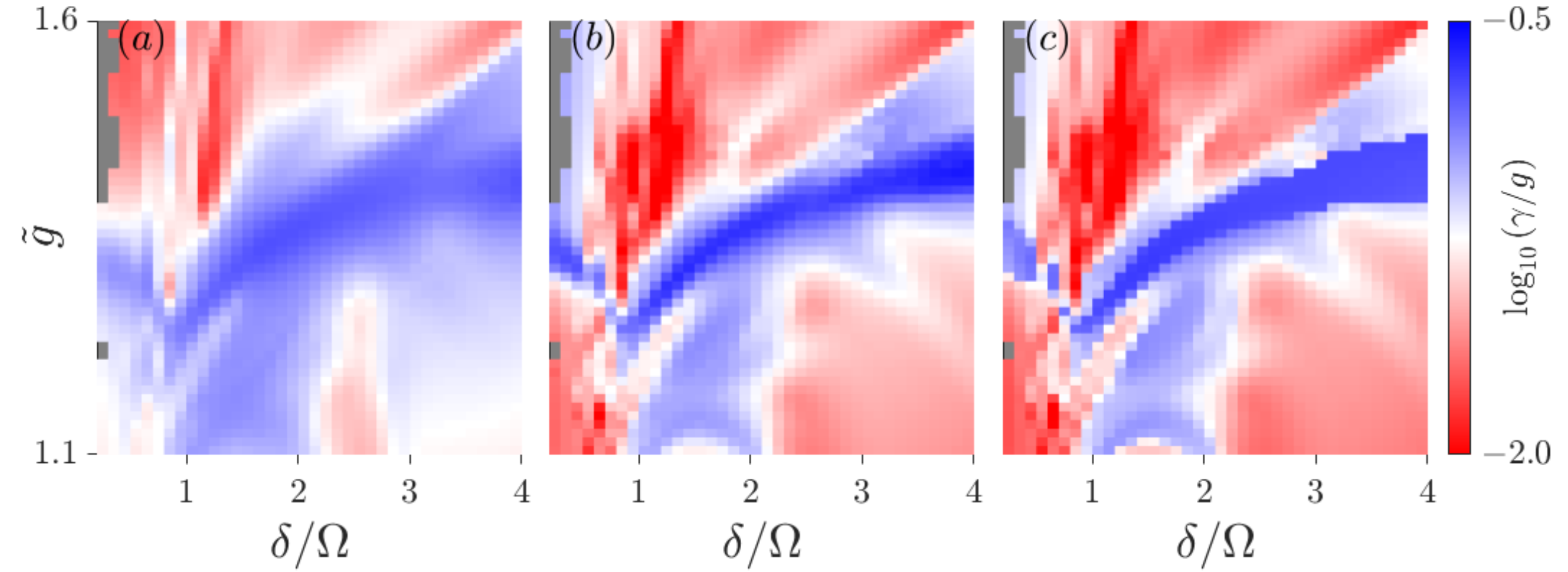}
    \caption{Comparison of relaxation timescales for the quantum dynamics with $N=600$. Panels (a) and (b) correspond to timescales $\gamma$ obtained through fits of the entropy $S_{vN}$ and the variance $\langle(\Delta\hat{S}^z)^2\rangle$, respectively, to the exponential function $A(1-e^{-\gamma t})$; panel (b) is identical to Fig.~4(d) of the main text. Panel (c) corresponds to the timescale $\gamma_{logistic}$ obtained through a fit to a shifted logistic function (see text).}
    \label{fig:SM_timescales}
\end{figure}

In Fig.~\ref{fig:SM_timescales}, we plot various notions of the relaxation timescales for the dynamics, including the value of $\gamma$ obtained by an exponential fit to the spin variance, which is plotted in Fig.~4(d) of the main text. Despite the complex nature of the dynamics over the wide range of behaviors we observe, we find that these timescales – obtained by fits to relatively simple functions – are generally consistent with each other, and thus serve as reliably proxies in systematically comparing the dynamics over the considered parameter space.

\begin{figure}[tb]
    \includegraphics[width=12cm]{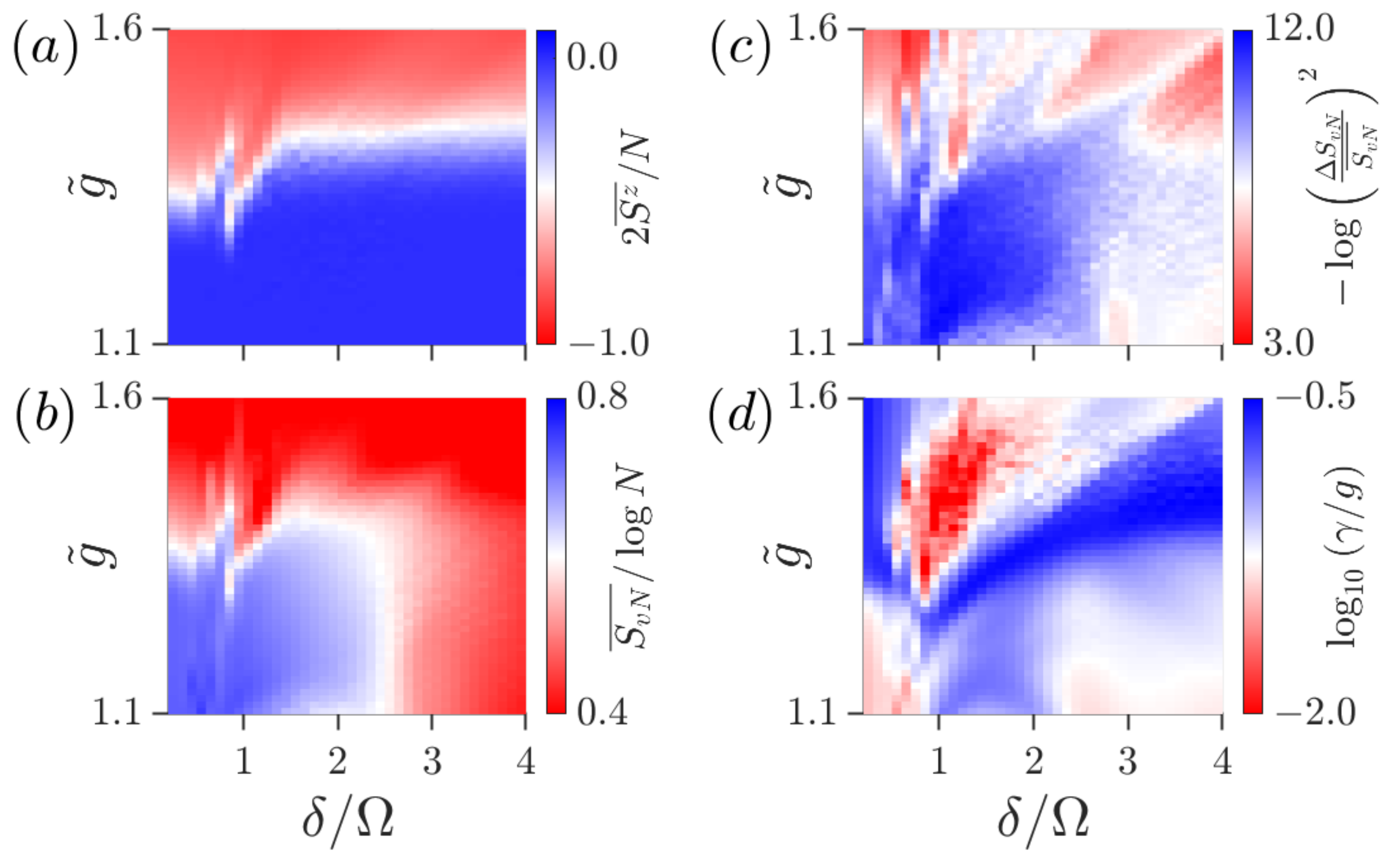}
    \includegraphics[width=12cm]{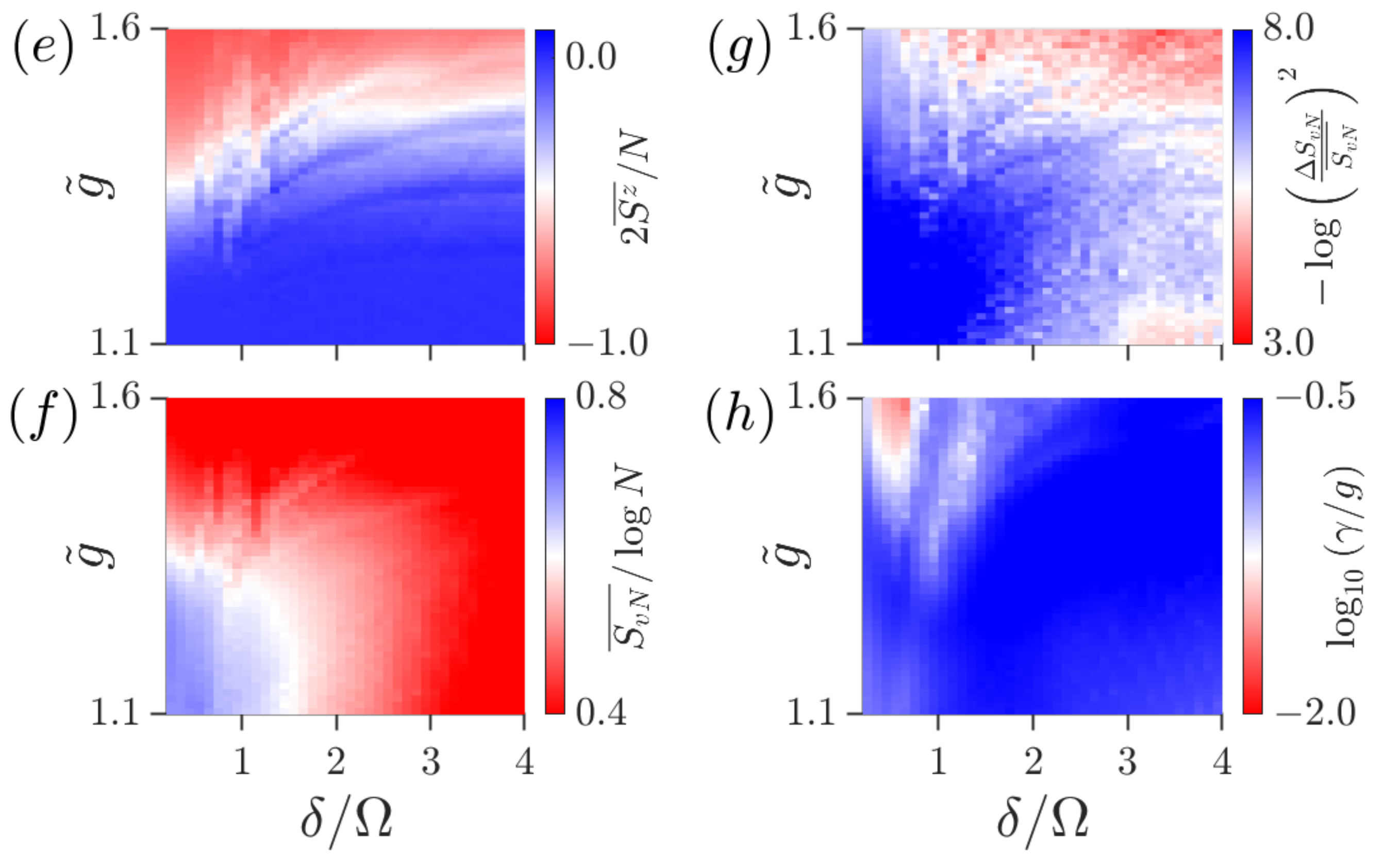}
    \caption{Comparison of quantum dynamics for $N=200$ (a-d) and $N=40$ (e-h). Panels (a)/(e) corresponds to the average magnetization, $\overline{ S^z } = (1/T)\int_{t_0}^{T+t_0} ~dt \langle\hat{S}^z(t)\rangle$. (b)/(f) Time-averaged entanglement entropy $\overline{ S_{\mathrm{vN}} } = (1/T)\int_{t_0}^{T+t_0} ~dt S_{\mathrm{vN}}(t)$ for $S_{\mathrm{vN}} = -\mathrm{Tr}[\hat{\rho}_s \mathrm{log}(\hat{\rho}_s)]$ where $\hat{\rho}_s$ is the reduced density matrix of the spins, analogous to Fig.~4(b) of the main manuscript. (c)/(g) Temporal fluctuations of entanglement entropy, $(\Delta S_{\mathrm{vN}})^2 = (1/T)\int_{t_0}^{T+t_0} ~dt [S_{\mathrm{vN}}(t) - \overline{ S_{\mathrm{vN}} }]^2$, normalized by $\overline{S_{\mathrm{vN}}}$, analogous to Fig.~4(c) of the main manuscript. (d)/(h) Relaxation timescale $\gamma$ of the dynamics. For panel (d), this is obtained by an empirical fit to $\langle (\Delta\hat{S}_z)^2 \rangle \propto (1 - e^{-\gamma t})$, analogous to Fig.~4(d) of the main manuscript; for panel (h), this is obtained by an empirical fit to $S_{\mathrm{vN}} \propto (1 - e^{-\gamma t})$. Time averages in (a)-(c), (e)-(g) are computed over a window $t\in[t_0,T+t_0]$ for $gt_0, gT = 150$, so as to reduce transient effects in steady state estimates.}
    \label{fig:SM_scaling}
\end{figure}

Lastly, in Fig.~\ref{fig:SM_scaling}, we also provide results for $N=200$ and $N=40$, analogous to those for $N=600$ shown in Fig.~4 of the main manuscript, with the addition of the steady-state magnetization $\langle\hat{S}^z\rangle$. For $N=40$, the timescale $\gamma$ is extracted via an empirical fit of the entanglement entropy $S_{vN}(t) \propto (1 - e^{-\gamma t})$, as opposed to the spin fluctuations $\langle (\Delta\hat{S}^z)^2\rangle$, which is used for $N=200$ and $N=600$; the spin fluctuations for smaller systems exhibit noisy oscillations that prevent a consistent estimate of a relaxation timescale. While signatures of the DPT are still evident in the single-body observables for systems as small as $N=40$, features of both the DPT and classical chaos are overshadowed by the associated quantum noise in this system. However, for $N=200$, which is within the capabilities of current trapped-ion platforms, signatures of classical chaos and the DPT feature much more prominently.

\section{Experimental realization in a trapped ion quantum simulator}
In the main text we discuss how the dynamical phases we report can be realized in an implemenation of the Dicke model in a $2$D trapped ion array formed in a Penning trap experiment \cite{Bollinger_2018,Cohn_2018}. Here, we support these comments with a brief comparison of the achievable coherent time-scales relative to decoherence and technical limitations. 

First, we establish the time-scales of the dynamics we are most interested in. To approach this problem concretely, we will focus on the limiting cases of the boson- and spin-dominated for which we can identify the relevant energy-scales from our prior analytic treatment in the mean-field limit. We expect quantum corrections to predominantly lead to, e.g., damping of oscillations due to the quantum noise. The resonant regime interpolates between the integrable spin and boson regimes, and thus should be suitably estimated from our results here. 

The key requirement of the experiment will be to distinguish the untrapped from trapped phase. In current experiments measurement of the phonon modes is difficult, and so we focus on the oscillations in the spin degree of freedom, $S_z$ (note that both $S_z$ and $X$ are always equally suitable observables to diagnose the DPT). In particular, we must be able to observe clear oscillations in $S_z$ when $\tilde{g} < \tilde{g}_{\mathrm{DPT}}$. This implies that the frequency of the oscillations must be large compared to relevant decoherence. In the $2$D Penning trap of Refs.~\cite{Bollinger_2018,Cohn_2018,Bohnet2016,Martin2017_OTOC,Arghavan_2017} the dominant source of decoherence is single-particle elastic dephasing of the spins at a rate $\Gamma_{\mathrm{el}}$. In comparison, the frequency scale of the relevant oscillations in $S_z$ in the boson-dominated regime is given by $\delta$\footnote{Fast oscillations in $S_z$ at frequency $\Omega$ are overlaid on top of the slower oscillation envelope. While it is possible, but perhaps workload intensive, to temporally resolve these in the current experiment, we further point out that constructing a reliable time-average of $S_z$ (and thus distinguishing the phases) does not require one to do so. Importantly, one must only ensure that data is not taken at, e.g., multiples of the period of these oscillations to avoid pathological aliasing/sampling problems in the data.}, whereas in the spin-dominated (LMG) regime the oscillation frequency is $\Omega$. Thus we require $\delta/\Gamma_{\mathrm{el}} \gtrsim 1$ for $\delta/\Omega \ll 1$ and $\Omega/\Gamma_{\mathrm{el}} \gtrsim 1$ for $\delta/\Omega \ll 1$.

Next, to make some approximate quantitative statements about the comparison of these time-scales let us, for simplicity, fix $\tilde{g} = 1$, as this is the characteristic scale of the DPT in both the spin- and boson-dominated regimes. Given a fixed ratio $\eta = \delta/\Omega$ we then have from the condition $\tilde{g} = 1$ that $\Omega = 2g/\sqrt{\eta}$ or $\delta = 2g\sqrt{\eta}$. Finally, this allows us to rewrite the condition on our oscillation frequency scales as
\begin{eqnarray}
    \frac{\delta}{\Gamma_{\mathrm{el}}} = \frac{2g\sqrt{\eta}}{\Gamma_{\mathrm{el}}} ~ \mathrm{for} ~ \eta \ll 1 , \\
    \frac{\Omega}{\Gamma_{\mathrm{el}}} = \frac{2g}{\sqrt{\eta}\Gamma_{\mathrm{el}}} ~ \mathrm{for} ~ \eta \gg 1 .
\end{eqnarray}

To connect to the experimental system we consider a pair of accessible spin-boson coupling strengths, $2g/(2\pi) = (1.88, 3.7)~\mathrm{kHz}$ \cite{KevinPrivateComm}. These values are similar to that reported in Ref.~\cite{Bollinger_2018}. The decoherence rate $\Gamma_{\mathrm{el}}$ is correlated with the spin-boson coupling strength, as it arises due to spontaneous emission from the applied off-resonant ODF lasers \cite{Bohnet2016}, and for the above values of $g$ we correspondingly have $\Gamma_{\mathrm{el}} \approx (350, 550)~\mathrm{s}^{-1}$. Thus, from an example calculation with $2g/(2\pi) = 3.7~\mathrm{kHz}$ we find $\delta/\Gamma_{\mathrm{el}} = \Omega/\Gamma_{\mathrm{el}} \approx 13$ at $\eta = 0.1$ and $\eta = 10$ respectively. Both values demonstrate that, in principle, the dynamical phases should be accessible across the full range of $\eta$ in the current trapped ion simulator.

\begin{figure}[b!]
    \includegraphics[width=8cm]{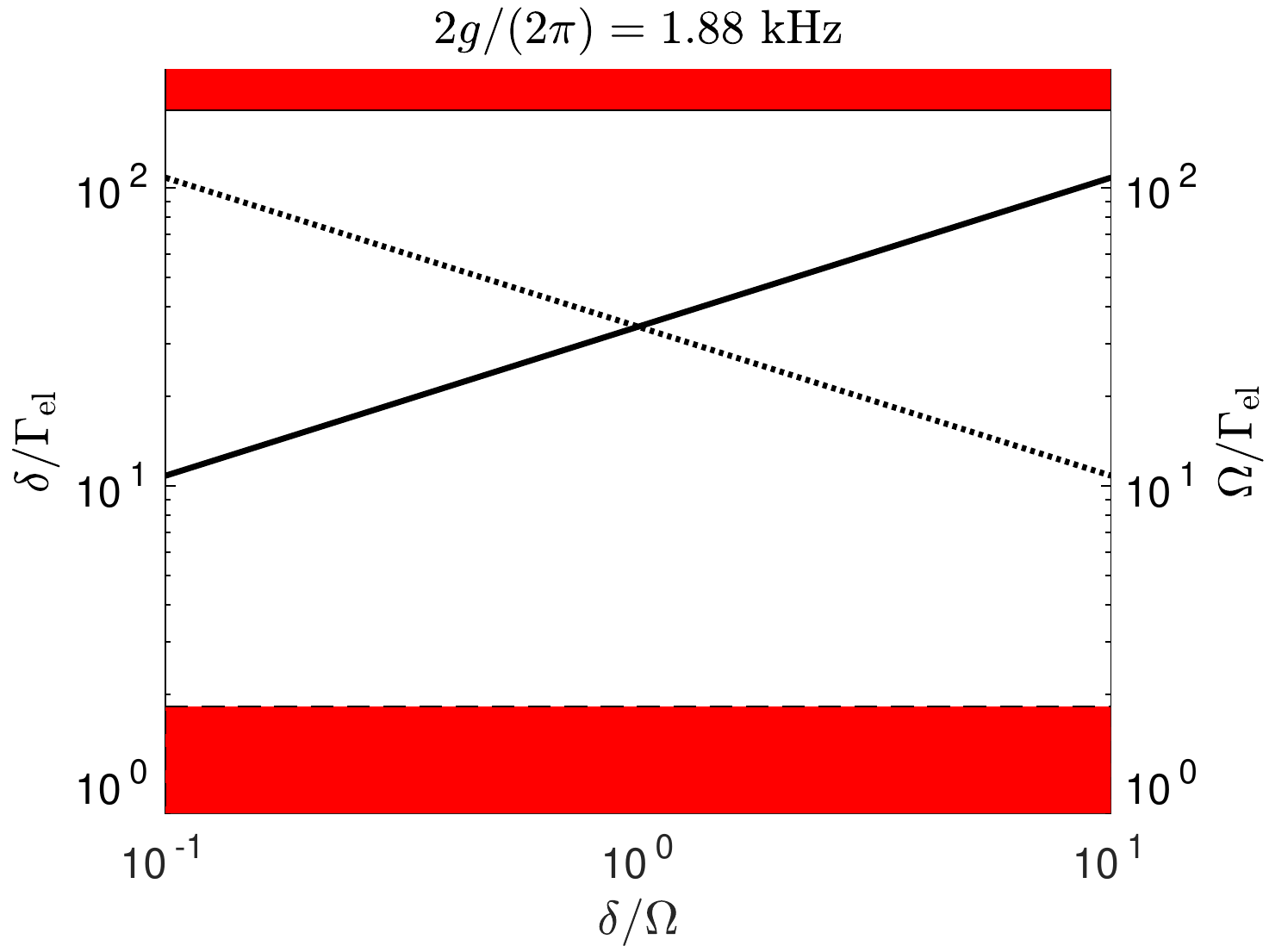}
    \includegraphics[width=8cm]{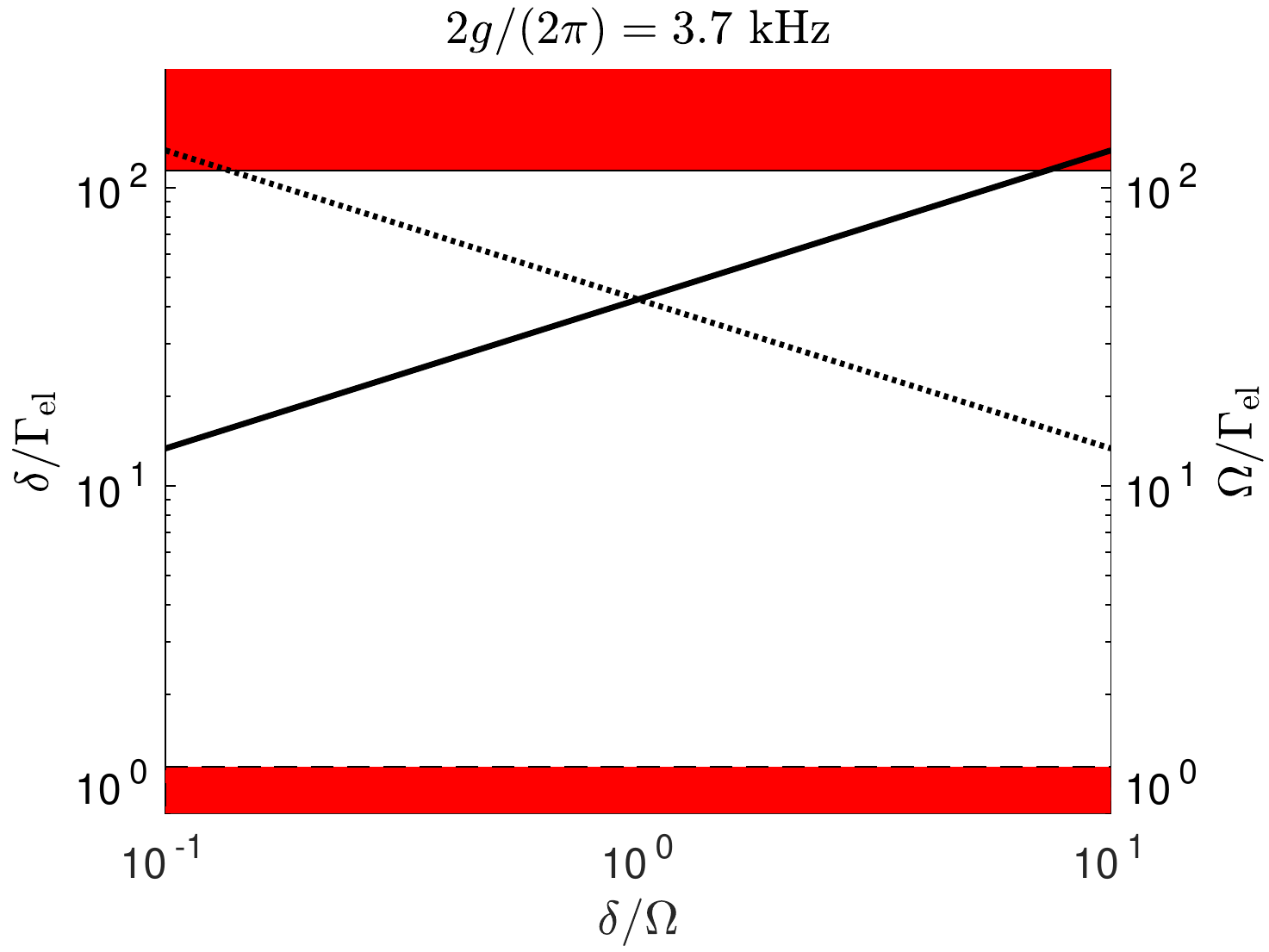}
    \caption{Comparison of coherent and dissipative time-scales for experimental system based on Ref.~\cite{Bollinger_2018}. Solid lines indicate $\delta/\Gamma_{\mathrm{el}}$ for boson-dominated regime ($\delta/\Omega \ll 1$), while dotted lines indicate $\Omega/\Gamma_{\mathrm{el}}$ for spin-dominated regime ($\delta/\Omega \gg 1$). Red shaded areas are relevant for comparison with $\delta/\Gamma_{\mathrm{el}}$: Upper red shaded area indicates detunings that may introduce couplings to other normal modes of the crystal, $\delta/(2\pi) > 10~\mathrm{kHz}$, while lower shaded area indicates detunings below the limit imposed by COM fluctuations $\delta/(2\pi) < 50~\mathrm{Hz}$.}
    \label{fig:FigSM_experiment}
\end{figure}

A more systematic and careful study is presented in Fig.~\ref{fig:FigSM_experiment}, where we also indicate the role of other technical limitations. In particular, the available range of detunings $\delta$ is limited by: i) coupling to other normal modes of the crystal and ii) fluctuations of the COM frequency related to impurity ions in the crystal and thermal occupation of the in-plane motional modes \cite{KevinPrivateComm}. The former places an approximate limit of $\delta/(2\pi) \lesssim 10~\mathrm{kHz}$ while the latter bounds the detuning from below as $\delta/(2\pi) \gtrsim 50~\mathrm{Hz}$. One might expect COM fluctuations could be impactful in regimes of $\eta \ll 1$ where small $\delta$ is favoured, while coupling to other modes will be relevant for $\eta \gg 1$ where $\delta$ is typically larger. Examining the predicted frequency scales in Fig.~\ref{fig:FigSM_experiment}, we find that COM fluctuations should not be an issue but coupling to other modes can become problematic for $\eta \gg 1$ if $g$ is set too large. In particular, one should favour $2g/(2\pi) = (1.88)~\mathrm{kHz}$ to ensure $\delta/(2\pi) < 10~\mathrm{kHz}$ at the largest values of $\eta$. For completeness, the transverse field takes values in the range $0.5~\mathrm{kHz} \lesssim \Omega/(2\pi) \lesssim 12~\mathrm{kHz}$ in both plots, which should be readily accessible.

\end{document}